\documentclass[12pt]{article}
\usepackage{amssymb}
\usepackage{latexsym}
\usepackage{graphicx}

\renewcommand{\thefootnote}{\fnsymbol{footnote}}

\def\appendix#1{
  \addtocounter{section}{1}
  \setcounter{equation}{0}
  \renewcommand{\thesection}{\Alph{section}}
  \section*{Appendix}
  \addcontentsline{toc}{section}{Appendix \thesection\ \ \ #1}
  }

\newcommand{\newsection}{    
\setcounter{equation}{0}
\section}

\newcommand{\eq}[1]{Eq.~(\ref{#1})}

\def\bea{\begin{eqnarray}}
\def\eea{\end{eqnarray}}

\def\be{\begin{equation}}
\def\ee{\end{equation}}

\def \bi{\bibitem}

\def\e9{\mbox{${E_9}$}}

\def\id{\protect{{1 \kern-.28em {\rm l}}}}

\begin{document}

\setcounter{page}{1}
\renewcommand{\thefootnote}{\arabic{footnote}}
\setcounter{footnote}{0}

\begin{titlepage}
\begin{flushright}
CU-TP-1003\\
\end{flushright}
\vspace{1cm}

\begin{center}
{\LARGE The Stability of\\
Noncommutative Scalar Solitons}

\vspace{1.1cm}
{\large{Mark G. Jackson}\footnote{
E-mail: markj@phys.columbia.edu} }\\

\vspace{18pt}

{\it Department of Physics}

{\it Columbia University}

{\it  New York City, NY 10027}
\\
\end{center}
\vskip 0.6 cm

\begin{abstract}
We determine the stability conditions for a radially symmetric 
noncommutative scalar soliton at finite noncommutivity parameter $\theta$.  
We find an intriguing relationship between the stability and existence 
conditions for all level-1 solutions, in that they all have 
nearly-vanishing stability eigenvalues at critical $\theta m^2$.  
The stability or non-stability of the system may then be determined 
entirely by the $\phi^3$ coefficient in the potential.
For higher-level solutions we find an ambiguity in extrapolating solutions 
to finite $\theta$ which prevents us from making any general statements.  
For these stability may be determined by comparing the fluctuation 
eigenvalues to critical values which we calculate.
\end{abstract}
\end{titlepage}
\newsection{Introduction}
\paragraph{}
There has recently been much interest, particularly from string theorists, in 
noncommutative geometry \cite{st1} \cite{st2}.  Part of this 
interest has focused on the construction of noncommutative scalar 
solitons, first addressed by Gopakumar {\it et al.} \cite{har}.  Explicit 
constructions were carried out for $\theta = \infty$ using 
an isomorphism between the star product and the simple harmonic oscillator 
(SHO) basis.  These solutions 
were then extrapolated to large but finite $\theta$ using the perturbed 
equations of motion, and some specific examples of solutions were 
constructed.  Classical stability was addressed only as an 
order-of-magnitude 
argument:  a stable solution at large $\theta$ changes mass eigenvalues by $\mathcal O (1 / \theta)$, allowing 
one to conclude that angular fluctuations (in the form of $U(\infty)$ 
transformations) give rise to instability in some solutions while still 
ensuring stability against radial fluctuations.  Existence of the 
solitons was then more thoroughly addressed in \cite{zhou} and \cite{dur}, where it was 
discovered that a potential bounded from below produces a critical 
$\theta$ below which there exists no solution; an unbounded potential will always 
yield a solution for any nonzero $\theta$.  This creates a need for  
detailed stability analysis at finite $\theta$.  In this paper we report progress in 
this direction.
\paragraph{} The paper is structured as follows.  In section 2
we review the construction of noncommutative solitons.  In section 3 we 
present the tools necessary to determine radial stability of a solution.  In section 
4 we generalize on the stability behavior for all theories.  We find that 
at the last point of existence, the critical value $(\theta m^2 )_c$, all 
level-1 solutions have almost neutral stability in that their eigenvalue is 
virtually identical to a critical eigenvalue which we calculate.  The small
deviation away from this critical eigenvalue is a product entirely of the 
corrections to the vacuum solution and thus to leading order is a function of the 
$\phi^3$ coefficient in the potential.  For a negative coefficient 
(those found in typical false-vacuum solutions) the solution will be 
unstable, whereas for a positive coefficient it will be stable.  For 
higher-level solutions we find an ambiguity in extrapolating solutions to 
finite $\theta$ which prevents us from making any general statements.  
For these stability may be determined by comparing the fluctuation 
eigenvalues to critical values which we calculate.
\newsection{Construction of Noncommutative Solitons}
\paragraph{} One begins with a field theory of a single scalar $\phi$ in 
$(2+1)$ dimensions with noncommutativity in the (complex) spatial directions $z$ 
and ${\bar z}$,
\begin{equation}
E = \frac{1}{g^2} \int d^2 z  (\partial _z \phi \partial _{\bar z} \phi + 
V(\phi) ),
\end{equation}
where $d^2 z = dx \ dy$.  Fields in this non-local action are multiplied using 
the Moyal star product
\begin{equation}
(A \star B) (z, {\bar z}) = e^{ \frac{\theta}{2} ( \partial _z \partial 
_{{\bar z}'} - \partial_{z'} \partial_{{\bar z}} )} A(z, {\bar z}) B(z', {\bar 
z}') | _{z = z'}.
\end{equation}
The most convenient way to obtain solutions to this theory is taking the 
$\theta \rightarrow \infty$ limit, rescaling $z \rightarrow z 
\sqrt{\theta}$, ${\bar z} \rightarrow {\bar z} \sqrt{\theta}$.  The kinetic 
term is then negligible and we can focus upon solutions to $V'(\phi)=0$.  
As $\phi$ is now an operator this may appear difficult, but it becomes 
greatly simplified by using the procedure outlined in \cite{har} whereupon 
radially symmetric $\phi(r)$ can be written in terms of projection operators in the SHO basis,
\begin{eqnarray}
\label{expr}
\phi(r) &=& \sum_{n=0} ^\infty c_n | n \rangle \langle n|, \\
&\equiv& \sum_{n=0} ^\infty c_n \phi_n (r).
\end{eqnarray}
so $(\phi \star \phi)(r) = \phi(r)$.  It is then simple to see 
we can write all radially symmetric solutions to the equation 
of motion in the form of \eq{expr},
with $c_n$ satisfying an equation of motion.  Angular transformations may then be 
performed by a $U(\infty)$ transformation, which is an exact symmetry in 
the $\theta = \infty$ limit.  The moduli 
space of all solutions in this limit can then be 
divided into unique {\it levels}.  For example, a ``level 2" solution can be 
obtained by acting with a $U(N)/U(N-2), \ N \rightarrow \infty$ symmetry on
\begin{equation}
\phi = c( \phi_0 + \phi_1).
\end{equation}
Stability 
determination is easy: if $c$ is a 
minimum (local or global), the solution is stable.  If it is a 
maximum, it is unstable.  
\paragraph{}
Upon making $\theta$ large but finite, one introduces the kinetic term
\begin{equation}
\label{kin}
K = \frac{2 \pi}{g^2} {\rm Tr} \ [a, \phi] [ \phi, a^\dagger].
\end{equation}
This will now break 
the $U(\infty)$ symmetry and change the equation of motion.  Stability determination 
requires examining the effect of angular fluctuations and radial fluctuations.  
Level 1 solution stability was considered in the following terms by \cite{har}.  
Angular fluctuations may be introduced in the form of fluctuations 
of $U$ which connect $\phi_0$ to $\phi_m, m \neq 0$.  These force the
solution to decay to $\phi_0$.  Now consider radial fluctuations for this solution.  Since we expect 
these to change by only $\mathcal O(1/\theta)$, it is stable or 
unstable as before.
\paragraph{} There is, however, a need for general stability analysis, 
valid at any finite $\theta$.  Zhou \cite{zhou} and Durhuus {\it et al.} 
\cite{dur} have analyzed the
conditions which allow soliton solutions at finite $\theta$.   It is the 
stability determination at finite $\theta$ that we present here.  We 
consider angular fluctuations by using a kind of Bogomolnyi bound on 
$\mathcal O(1/\theta)$ kinetic fluctuations as explained by Gopakumar and Headrick \cite{talk},
\begin{eqnarray}
\delta K (\phi) &=& \frac{2 \pi}{g^2} \left[ \left( {\rm Tr} \ \phi \right) + {2 \sum} _{|\alpha \rangle \in 
\mathcal H 
_\perp, |i \rangle \in \mathcal H _{\parallel}} | \langle \alpha | a | i 
\rangle |^2 \right], \\
&\geq& \frac{2 \pi}{g^2} {\rm Tr} \ \phi
\end{eqnarray}
where $\mathcal H _{\parallel}$ is the Hilbert space of projection 
operators 
corresponding to $\{ \phi_0, \ldots, \phi_{n-1} \}$ and $\mathcal H _{\perp}$ 
is the Hilbert space of all others.
This shows that radially symmetric solutions saturate the bound, and it is 
these solutions we focus upon here.  One can then use the procedure presented 
in the next section to determine radial fluctuation stabilty at finite $\theta$. 
We should comment on how much we can trust our conclusions, if we 
can only trust angular stability to $\mathcal O (1/\theta)$.  First,
angular fluctuations typically cost more energy and should be less 
significant than radial fluctuations at any $\theta$.  Second, since it was 
shown in \cite{zhou} that $(\theta m^2)_c \geq 13.92$ for $\phi^4$ theories, 
$1/ \theta m^2$ is likely still a 
somewhat small parameter.  
\newsection{Radial Fluctuation Stability}
\subsection{Determination of Radial Fluctuation Eigenvalues}
\paragraph{}
One of the nice features of the SHO basis expansion of 
the soliton solutions is that we have an explicit basis $\{ c_n \}$ from 
which to compute energy $E ( \{ c_n \} )$, equations of motion $\partial 
E / \partial c_n = 0$, and fluctuation matrix $\partial ^2 E / \partial c_n 
\partial c_m$.  The latter must have all positive eigenvalues for a stable 
solution.
\paragraph{} 
The energy is given as 
\begin{equation}
E( \{ c_n \} ) = \sum _{n=0} ^{\infty} [ (2n+1) c_n ^2 - 2(n+1) c_{n+1} 
c_n + \theta V(c_n) ]
\end{equation}
and thus the stability matrix is
\begin{equation}
\frac{\partial ^2 E}{\partial c_n \partial c_m} =[\theta V''(c_n) + 
2(2n+1)] 
\delta _{nm} - 2(n+1) (\delta _{n,m+1} + \delta _{n,m-1})
\end{equation}
which is more clearly written as
\begin{equation}
\left( \begin{array}{cccc}
\theta V''(c_0) + 2 & -2 & &  \\
-2 & \theta V''(c_1) + 6 & -4 &  \\
& -4 & \theta V''(c_2) + 10 & -6  \\
& & -6 &   \ddots 
\end{array} \right).
\end{equation}
We may determine the 
eigenvalues by putting this in the upper-diagonal form
\begin{equation}
\left( \begin{array}{cccc}
\theta V''(c_0) + 2 & -2 & & \\
 & \theta V''(c_1) + 6 - \frac{2^2}{\theta V''(c_0) + 2}& -4 &  \\
&  & \theta V''(c_2) + 10 - \frac{4^2}{\theta V''(c_1) + 6 - 
\frac{2^2}{\theta V''(c_0) + 2}} & -6 \\
& &  &   \ddots 
\end{array} \right)
\end{equation}
from which we can read off the eigenvalues
\begin{equation}
\label{eigs}
\lambda_{n} = \theta V''(c_n) + 2 (2n+1) - \frac{(2n)^2}{\lambda 
_{n-1}}.
\end{equation}
\subsection{Restrictions on Eigenvalues}
\paragraph{} It seems we have an obvious method to determine the eigenvalues: 
armed with 
the $\{ c_n \}$ of a solution, compute $\lambda _0$ and see if it is 
positive.  If it is, compute $\lambda _1$ and see if it is, and so forth.  
But we need to 
determine the positivity of {\it all} $\lambda_n$, not just the 
first few; how can we determine all of them?  We do this by assuming 
$\lambda _n>0$, and then considering 
what bound is placed on $\lambda_n$ that ensures $\lambda _{n+1}$ is 
positive.  This is done by using the recurrence relation
\begin{equation}
\lambda_n = \frac{(2(n+1))^2}{\theta V''(c_{n+1}) + 2(2n+3) - \lambda 
_{n+1}},
\end{equation}
then setting $\lambda_{n+1} = 0$ to get
\begin{equation}
\lambda_n > \frac{(2(n+1))^2}{\theta V''(c_{n+1}) + 2(2n+3)}.
\end{equation}
Note that this is more restrictive than simply $\lambda _n > 0$.  Now 
assuming that $\lambda _{n+1} >0$, what bound on $\lambda_n$ now ensures 
that $\lambda_{n+2}$ is 
positive?  Using the relation between $\lambda_{n+1}$ and $\lambda_{n+2}$ 
and then setting the latter to 0,
\begin{equation}
\lambda_n > \frac{(2(n+1))^2}{\theta V''(c_{n+1}) + 
2(2n+3)-\frac{(2(n+2))^2}{\theta 
V''(c_{n+2}) + 2(2n+5)}}.
\end{equation}
Similarly, this is again more restrictive than the previous condition.  
Thus 
the condition placed on $\lambda_n$ so that {\it all} $\lambda$ after it 
are 
positive is the obvious limit,
\begin{eqnarray}
\lambda_n &>& \frac{(2(n+1))^2}{\theta V''(c_{n+1}) + 
2(2n+3)-\frac{(2(n+2))^2}{\theta 
V''(c_{n+2}) + 2(2n+5) - \cdots}} \\
&\equiv&  \lambda_n ^{crit} \\
&=&  \frac{(2(n+1))^2}{\theta V''(c_{n+1}) + 2(2n+3) - \lambda 
_{n+1}^{crit}}.
\end{eqnarray}
This appears to be of little practical use since we need to 
know 
all $\{ c_n \}$ explicitly.  But since we are only interested in finite-energy solutions we 
know all $c_n \rightarrow 0$ for large $n$ and we can get away with 
$V''(c_n) \approx m^2 + \mathcal O (c_n)$.  Thus our strategy is this:  compute the first few $\lambda_n$ 
by hand using the $\{ c_n \}$, making sure each is positive.  When the remaining 
$c_n$ are small ($c_n=0$ at $\theta = \infty$), 
simply check whether 
the last $\lambda_n$ computed by hand exceeds $\lambda _n ^{crit}$ plus a 
small correction due to the small $c_n$.  If 
it does, the solution is stable.  Since we already have explicit 
expressions for $\lambda_n$, the first part of this is already done; all 
that remains is to calculate explicit expressions for $\lambda_n ^{crit}$ for $V''(c_n) = m^2$ and work 
out the correction terms.  This can be done by considering 
a general potential
\begin{equation}
V(\phi) = \frac{1}{2} m^2 \phi^2 + \frac{1}{3} a \phi^3 + \cdots
\end{equation}
allowing us to approximate $V''(c_n) \approx m^2 + 2a c_n$.  If $a=0$, we 
use the next non-vanishing term.  We then 
compute $\lambda _n (\theta m^2)$ and the perturbation $\Delta \lambda _n 
^{crit} (\theta m^2)$:
\begin{eqnarray}
\label{crit}
\lambda _n ^{crit} (\theta m^2) &=& \frac{(2(n+1))^2}{\theta m^2 + 
2(2n+3)-\lambda^{crit} 
_{n+1}} ,\label{lcrit} \\ 
\Delta \lambda _n ^{crit} &=& \left( 2 \theta a c_{n+1} \frac{\partial}{\partial 
(\theta m^2)} + \Delta \lambda _{n+1} ^{crit} \frac{\partial}{\partial \lambda 
_{n+1} ^{crit}} \right) \lambda _n ^{crit} \\
&=& - \left( \frac{ \lambda _n 
\label{cor}
^{crit} }{2(n+1)} \right) ^2 ( 2 \theta a c_{n+1} - \Delta \lambda _{n+1} 
^{crit} ).
\end{eqnarray}
\subsection{Calculation of $\lambda _n ^{crit} (\theta m^2)$}
\paragraph{}
We now use \eq{crit} to derive an explicit formula for $\lambda _n ^{crit} (\theta m^2)$.  
At first glance this appears to be just our stability matrix eigenvalues 
if all $V''(c_n) = m^2$, and thus $\lambda _0 ^{crit} = \theta m^2 + 2$, 
etc.  In fact this is not the case.  The reason is that this has two roots, determined by the method by which we compute the 
$\lambda _n$.  Computing in the order $\lambda_0, \lambda_1, \ldots, 
\lambda_{n-1}$ 
selects one root, 
whereas computing $\lambda_\infty, \ldots, \lambda _{n+1}$ selects the 
other.  Since our $\lambda_n ^{crit}$ clearly 
depends upon the higher $\lambda$'s as just explained, we need to 
choose this latter root.  In the appendix we show that
\begin{equation}
\lambda _0 ^{crit} (\theta m^2)= \theta m^2+2-\frac{2}{e^{\theta m^2/2} \Gamma(0,\frac{\theta m^2}{2})}
\end{equation}
where $\Gamma(a,z)$ is the incomplete Gamma function.  This can be used 
with the recursion relation \eq{crit} to easily calculate the next terms.  We plot 
$\lambda_0 ^{crit} (\theta m^2)$ in Figure 1; the qualitative behavior for 
all levels is similar.
\begin{figure}[t]
\begin{center}
\includegraphics[scale=0.6]{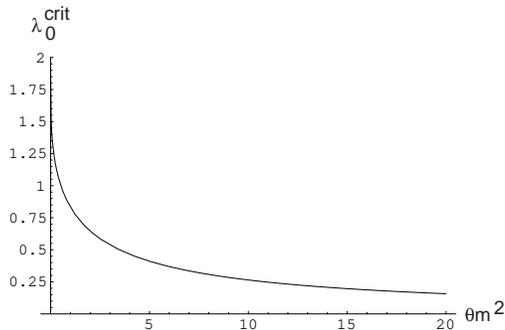}
\caption{Critical eigenvalue $\lambda _0 ^{crit}$ as a function of $\theta m^2$.}
\end{center}
\end{figure}
\paragraph{} Two important results are immediately obvious from the 
solution.  
First, as $\theta m^2 \rightarrow \infty$ we 
see $\lambda _n ^{crit} \rightarrow 0$, as we would expect from the 
interpretation that the eigenvalues
decouple and must simply be positive ``on their own".  Second, the 
critical 
eigenvalues aren't real for $m^2<0$, so, as expected, there are no stable 
solutions unless there is a stable true vacuum.
\paragraph{} We then consider the corrections $\Delta \lambda _n ^{crit}$,
\begin{equation}
\Delta \lambda _n ^{crit} = - 2 \theta a \left( \frac{ \lambda _n ^{crit} }{2(n+1)} \right) ^2 
\left( c_{n+1} +  
\left( \frac{ \lambda _{n+1} ^{crit} }{2(n+2)} \right) ^2 \left( c_{n+2}  
+ 
\cdots \right) \right).
\end{equation}
Since we know $\lambda _n ^{crit} \leq 2(n+1),$
\begin{eqnarray}
| \Delta \lambda _n ^{crit} | &\leq& | 2 \theta a (c_{n+1} + c_{n+2} + \cdots 
)| \\
&\approx & \left| \frac{2 \theta a}{m^2} \left( V'(c_{n+1}) + V'(c_{n+2}) + 
\cdots \right) 
\right| \label{veq}
\end{eqnarray}
The convergence of \eq{veq} was proved by \cite{zhou} so this correction 
is well-defined.  In fact the corrections are very small in most circumstances.
\section{General Stability Analysis}
\paragraph{} Naturally one wonders what general statements about a 
solution's stability are possible without checking the eigenvalues numerically.  
\subsection{Stability of the Gaussian} Let us focus on the Gaussian case to begin with; this 
is the unique level-1 solution at $\theta = \infty$ which is stable with respect to 
angular fluctuations.  The solution requires $(\epsilon \equiv 1/ \theta m^2 )$
\begin{eqnarray}
\label{bc1}
\lim_{\epsilon \rightarrow 0} c_0 &\rightarrow& c,\\
\label{bc2}
\lim_{\epsilon \rightarrow 0} c_n &\rightarrow& 0
\end{eqnarray}
uniformly in $n$ for $n \geq 1$.  Let us quickly review the procedure to 
construct solutions at finite $\theta$, given in \cite{har}.  The perturbed equation 
of motion is
\begin{equation}
(-\epsilon \partial ^2 + 1) \phi = b \phi_0
\end{equation}
for some scale parameter $b$.  This can be solved in Fourier-space as
\begin{equation}
{\tilde \phi}(k) = b \frac{{\tilde \phi} _0 (k)}{1 + \epsilon k^2} .
\end{equation}
Using expressions for ${\tilde \phi}(k)$ 
in terms of Laguerre polynomials $L_n (x)$ we can then calculate the solution from
\begin{equation}
c_n(\epsilon) = b \int _0 ^\infty \frac{e^{-x} L_n (x)}{1 + 2 \epsilon x} \ 
dx .
\end{equation}
The normalization of $L_n(x)$ is such that $\int _0 ^\infty e^{-x} L_n(x) 
L_m(x) dx = \delta_{mn}$.  We determine $b$ by matching boundary conditions in \eq{bc1} and \eq{bc2} to get
\begin{equation}
V'(b F(\epsilon)) = m^2 b(F(\epsilon)-1), \hspace{0.5in} F(\epsilon) = \int _0 ^\infty 
\frac{e^{-x}}{1+2 \epsilon x} \ dx.
\end{equation}
As explained in more 
detail by \cite{zhou} and \cite{dur}, the soliton ceases to exist when there is no such 
$b$.  Just as finding solutions to the equation of motion at $\theta = 
\infty$ meant finding the real roots of $V'$, we can think of this as
finding the real roots of
\begin{equation}
V'_{eff} (bF(\epsilon)) \equiv V'(b F(\epsilon)) - m^2 b(F(\epsilon)-1).
\end{equation}
\paragraph{} At $\epsilon_c \equiv 1/(\theta m^2)_c$ the solution must also satisfy $\partial 
V'_{eff} / \partial b = 0$ (see Figure 2, for example), so
\begin{equation}
V''(b_c F(\epsilon_c)) = m^2 \left(1 - \frac{1}{F(\epsilon_c )} \right).
\end{equation}
Plugging this into our expression for $\lambda_0$ provides a simple answer for the 
eigenvalue at the last point of existence for the solution, completely 
independent of potential:
\begin{eqnarray}
\lambda _0 ((\theta m^2)_c) &=& \theta V''(c_0(\epsilon_c)) + 2 \\
&=& (\theta m^2)_c \left(1 - \frac{1}{F(\epsilon_c)} \right) +2 \\
&=& \lambda _0 ^{crit} ((\theta m^2)_c)
\end{eqnarray}
where we use the identity proved in the appendix.
\begin{figure}[t]
\begin{center}
\includegraphics[scale=0.6]{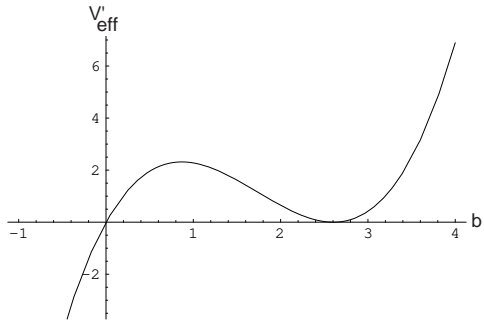}
\caption{At $(\theta m^2)_c$, $b$ will be a local minima of $V_{eff}$.}
\end{center}
\end{figure}
\paragraph{} Then our Gaussian stability comes down to only determining the 
sign of $\Delta \lambda _0 ^{crit}$.  If positive, it is slightly unstable; 
if negative, it is slightly stable.  Examining the form of $\Delta 
\lambda_0 ^{crit}$ in \eq{cor} we see this is entirely determined by the sign 
of the $\phi^3$ coefficient $a/3$, since it was shown in $\cite{zhou} 
\cite{dur}$ all $c_n>0$ if they come from $c_n=0$ at $\theta = \infty$.
\paragraph{} We provide an example of this using the false-vacuum potential given by 
\cite{zhou},
\begin{equation}
V(\phi) = 3 \phi^2 - \frac{5}{3} \phi^3 + \frac{1}{4} \phi^4.
\end{equation}
\begin{figure}[t]
\begin{center}
\includegraphics[scale=0.6]{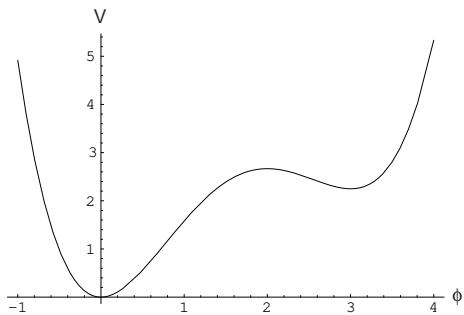}
\caption{A typical false-vacuum potential.}
\end{center}
\end{figure}
\paragraph{} This is graphed in Figure 3.  Choose the solution $\phi = 3 \phi_0$ at 
$\theta = \infty$, so that we begin with a stable solution.  It is easy to show 
that for this potential $(\theta m^2)_c = 
46.1488$.  From the procedure outlined this implies
\begin{eqnarray}
\lambda_0 ((\theta m^2)_c) &=& 0.0771,\\
\lambda_0 ^{crit} ((\theta m^2)_c) &=& 0.0771, \\
\Delta \lambda_0 ^{crit} ((\theta m^2)_c) &=& 0.0355.
\end{eqnarray}
In Figure 4 we plot $\lambda_0$ and $\lambda_0 ^{crit}$ near the critical 
point, seeing that indeed the solution has nearly neutral stability at its critical 
point.  When we then consider the effect of raising the critical 
eigenvalue by $\Delta \lambda_0 ^{crit} ((\theta m^2)_c) = 0.0355$, we 
know it is unstable.
\begin{figure}[t]
\begin{center}
\includegraphics[scale=0.6]{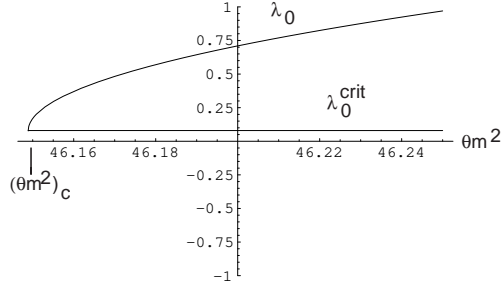}
\caption{$\lambda_0$ and $\lambda_0 ^{crit}$ near the critical point.}
\end{center}
\end{figure}
\subsection{Higher Levels}
\paragraph{}  Let us now attempt to extend this to the level-$n$ solution,
\begin{equation}
\phi = c(\phi_0 + \ldots + \phi_{n-1}).
\end{equation}
Now $\phi(x)$ will obey
\begin{equation}
(-\epsilon \partial^2 + 1) \phi(x) = \sum _{j = 0} ^{n-1} b_j \phi_j(x)
\end{equation}
and we can obtain coefficients from
\begin{equation}
\label{eom2}
c_i = \sum _{j=0} ^{n-1} b_j F_{ij} (\epsilon)
\end{equation}
where
\begin{equation}
F_{ij} (\epsilon) \equiv \int _0 ^\infty \frac{e^{-x}}{1+2\epsilon x} L_i (x) 
L_j(x) 
\ dx
\end{equation}
which is clearly symmetric under $i \leftrightarrow j$.  The equations of motion 
combined with boundary conditions are
\begin{equation}
\label{yeah}
\frac{1}{m^2} V'(c_i) = -b_i + \sum _{j=0} ^{n-1} b_j F_{ij}.
\hspace{0.5in} (i = 0, \ldots, n-1)
\end{equation}
Now rewrite this in the $\{ c_i \}$ basis using the substitution
\begin{equation}
b_i = {\bf F}_{ij} ^{-1} c_j
\end{equation}
where ${\bf F}_{ij} ^{-1}$ is the matrix inverse of ${\bf F}_{ij} \equiv \{ 
F_{ij} \}$.  These equations then need to be decoupled to obtain 
solutions for each $c_i$.  For $\theta = \infty$ this decoupling is trivial 
since $F_{ij} (\epsilon = 0) = \delta _{ij}$, producing $V'(c_i) = 0$ as 
expected.  But at finite $\theta$ the equations will include terms such as 
$V'(V'(c_i))$ and be of higher order.  This 
implies there will be an increase in the number of solutions.  It is then 
potentially ambiguous how to extrapolate 
$\theta = \infty$ solutions into finite $\theta$, and thus it is unclear how one could single 
out the solution which remains at lowest $\theta$.  To our knowledge, 
this solution ambiguity has not been discussed before.
\begin{figure}[t]
\begin{center}
\includegraphics[scale=0.65]{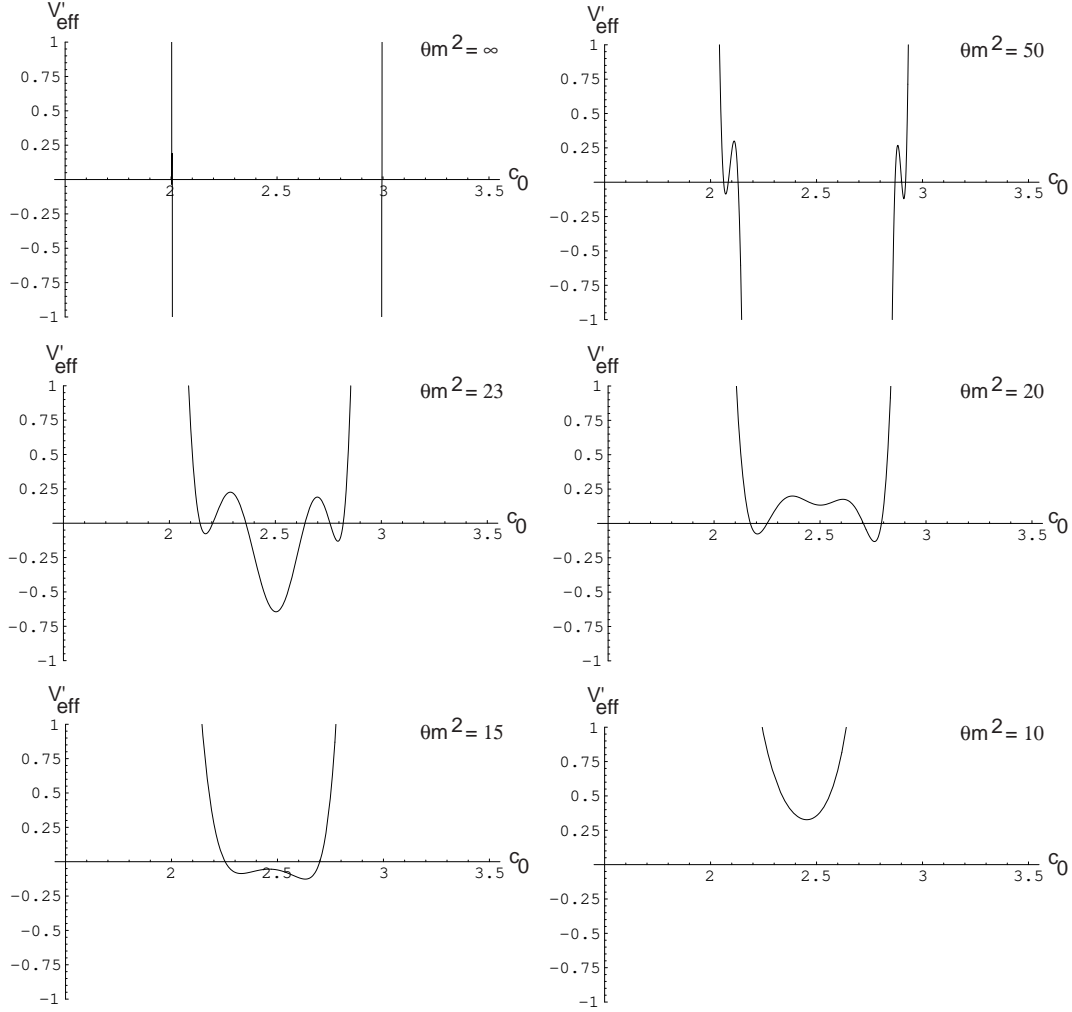}
\caption{$V'_{eff} (c_0)$ at several values of $\theta m^2$.}
\end{center}
\end{figure}
\paragraph{}
Let us demonstrate this concept explicitly in a level-2 solution.  \eq{yeah} 
becomes
\begin{eqnarray}
V'(c_0) &=& b_0 (F_{00} -1) + b_1 F_{01}, \\
V'(c_1) &=& b_0 F_{01} + b_1 (F_{11} -1 ).
\end{eqnarray}
Changing basis allows us to write these as
\begin{eqnarray}
V'(c_0) &=& \alpha c_0 + \beta c_1, \\
V'(c_1) &=& \beta c_0 + \gamma c_1.
\end{eqnarray}
where
\begin{equation}
\alpha = \frac{1}{\Delta} (\Delta - F_{11}), \hspace{0.4in} \beta = 
\frac{F_{01}}{\Delta}, \hspace{0.4in} \gamma = \frac{1}{\Delta} (\Delta - 
F_{00} ), \hspace{0.4in} \Delta \equiv F_{00} F_{11} - F_{01}^2.
\end{equation}
These are easily decoupled to get
\begin{eqnarray}
\frac{1}{m^2} V'\left( \frac{1}{\beta} \left( V'(c_1) - \gamma c_1 \right) \right) &=& 
\frac{\alpha}{\beta} \left( V'(c_1) - \gamma c_1 \right) + \beta c_1, \\
\label{graphed}
\frac{1}{m^2} V'\left( \frac{1}{\beta} \left( V'(c_0) - \alpha c_0 \right) \right) &=& 
\frac{\gamma}{\beta} \left( V'(c_0) - \alpha c_0 \right) + \beta c_0.
\end{eqnarray}
\paragraph{} We show in Figure 5 the solutions to \eq{graphed} at various 
values of $\theta$, using the same false vacuum potential as we did in the Gaussian 
case.  
Multiple solutions are clearly evident at 
finite $\theta$, each of which has a different $(\theta m^2)_c$.
\section{Conclusion}
\paragraph{} We have presented the conditions by which radial stability of 
a noncommutative scalar soliton may be determined.  One may explicitly 
calculate the stability eigenvalues of the solution and then compare these 
to critical eigenvalues which ensure the stability of the system.  For a 
level-1 solution, we found that the existence and stability of a system are 
closely intertwined, allowing one to make definite statements about the 
stability of a system at its critical existence point.  For level-$n$ 
solitons we do not believe such a general statement can be made, due to the
nonuniqueness of solutions at finite $\theta$.
\paragraph{}
There are many possible extensions of the analysis presented here.  The first would 
be a complete analysis of these extra level-$n$ solutions at finite 
$\theta$.  Another would be a comparison with the work by Aganagic {\it et al.} \cite{har2} in which unstable solitons
 in noncommutative gauge theory were 
addressed.  Also interesting would be to place this in the context of the large literature in which noncommutative 
methods have been applied to unstable D-branes and tachyonic condensation.  Yet another is that our methods only address classical stability and 
it would be very interesting to include quantum stability.
\section{Acknowledgements}
\paragraph{}
The author thanks D. Tong, B. Greene, D. Goldfeld, R. Gopakumar, and H. Peiris for helpful 
comments.  This work was supported by a GAANN Fellowship
from the United States Department of Education.
\section{Appendix: Equivalence of formulae}
\paragraph{} Here we present a proof for the equality of
\begin{equation}
\label{main}
x(1-1/F(x)) +2 = \lambda^{crit} _0 (x)
\end{equation}
where
\begin{eqnarray}
F(x) &=& \int _0 ^\infty \frac{e^{-u}}{1+2 u/x} \ du \\
&=& \frac{x}{2} e^{x/2} \Gamma (0, \frac{x}{2})
\end{eqnarray}
and
\begin{equation}
\lambda _n ^{crit} (x) = \frac{(2(n+1))^2}{x + 2(2n+3)-\lambda^{crit} 
_{n+1} (x)}.
\end{equation}
We can re-write \eq{main} as
\begin{eqnarray}
\label{main2}
x +1 - \frac{\lambda^{crit} _0 (2x)}{2} &=& \frac{x}{F(2x)} \\
x + 1 - \frac{1^2}{x+3-\frac{2^2}{x+5-\ldots}} &=& \frac{1}{e^x \ \Gamma(0,x)}
\end{eqnarray}
To prove this we use an identity found in Perron \cite{per} \S 82 formula (26)
\begin{equation}
\label{per}
x + \alpha + \beta + 1 - \frac{(\alpha + 1)(\beta + 1)}{x + \alpha + 
\beta + 3 - \frac{(\alpha + 2)(\beta + 2)}{x+\alpha + \beta + 5 - 
\ldots}} = \frac{\beta \int _0 ^\infty \frac{e^{-u} u ^{\beta-1}}{(x+u)^\alpha}\ 
du}{\int _0 ^\infty \frac{e^{-u} u ^{\beta}}{(x+u)^{\alpha+1}}\ 
du}
\end{equation}
which holds for $\beta \geq 0, \alpha \geq 0, x \geq 0$.  We note that
\begin{eqnarray}
\lim _{\beta \rightarrow 0} \beta \int _0 ^\infty e^{-u} u^{\beta-1} du &=& 
\lim _{\beta \rightarrow 0} \beta \ \Gamma (\beta) \\
&=& 1.
\end{eqnarray}
Using this we take the $\alpha = 0, \beta \rightarrow 0$ limit of \eq{per} 
to find
\begin{eqnarray}
x + 1 - \frac{1^2}{x+3-\frac{2^2}{x+5-\ldots}} &=& \frac{1}{\int _0 ^\infty 
\frac{e^{-u}}{x+u} \ du} \\
\label{main3}
&=& \frac{1}{e^x \ \Gamma(0,x)}.
\end{eqnarray}
We thank D. Goldfeld for providing us with this proof.

\end{document}